\documentclass[aps,twocolumn,showpacs,amsmath,amssymb,floatfix]{revtex4}
\usepackage{young}
\usepackage{graphicx}
\usepackage{color}
\usepackage[latin1]{inputenc} 
\def\nn{\nonumber}

\def \bc {\begin{center}}
\def \ec {\end{center}}
\def \bi {\begin{itemize}}
\def \ei {\end{itemize}}

\def \ba {\begin{array}}
\def \ea {\end{array}}

\def \bea {\begin{eqnarray}}
\def \eea {\end{eqnarray}}

\def \um {\frac{1}{2}}
\def\tr {\mathrm{tr}}
\def\cD {{\cal D}}

\def\DZ {\Delta_\mathrm{Z}}
\def\DS {\Delta_\mathrm{t}}
\def\ic {\mathrm{i}}

\newcommand{\la}{\langle}
\newcommand{\ra}{\rangle}

\begin{document}

\title{Hilbert space and ground state structure of bilayer quantum Hall systems at $\nu=2/\lambda$}

\author{M. Calixto, C. Pe\'on-Nieto and E. P\'erez-Romero}

\affiliation{Departamento de Matem\'atica Aplicada and Instituto ``Carlos I'' de F\'isica Te\'orica y Computacional,  Universidad de Granada,
Fuentenueva s/n, 18071 Granada, Spain}

\date{\today}

\begin{abstract}
We analyze the Hilbert space and ground state structure of bilayer quantum Hall (BLQH) systems at fractional filling factors $\nu=2/\lambda$ ($\lambda$ odd) and we also study the large 
$SU(4)$ isospin-$\lambda$ limit. The model Hamiltonian is an adaptation of the $\nu=2$ case [Z.F. Ezawa {\it et al.}, Phys. Rev. {B71} (2005) 125318] 
to the many-body situation (arbitrary $\lambda$ flux quanta per electron). 
The semiclassical regime and quantum phase diagram (in terms of layer distance, Zeeeman, tunneling, etc, control parameters) is obtained by using previously introduced Grassmannian $\mathbb{G}^4_{2}=U(4)/[U(2)\times U(2)]$ coherent states 
as variational states. The existence of three quantum phases (spin, canted and ppin) is common to any $\lambda$, but the phase transition points depend on $\lambda$, and the instance 
$\lambda=1$ is recovered as a particular case.  We also analyze the quantum case through a numerical diagonalization of the Hamiltonian 
and compare with the mean-field results, which give a good approximation in the spin and ppin phases but not in the canted phase, where we detect exactly $\lambda$ energy level 
crossings between the ground and first excited state for given values of the tunneling gap. An energy band structure at low and high interlayer tunneling (spin and ppin phases, respectively) 
also appears depending on angular momentum and layer population imbalance quantum numbers.

\end{abstract}


\maketitle

\section{Introduction}

A better analytical understanding of the Hilbert space and ground state structure of multicomponent fractional quantum Hall systems 
is needed to have a clear physical picture and to properly interpret the experimental data. In this article we make a contribution in this direction, studying the 
bilayer case at fractional values of the filling factor $\nu=2/\lambda$ ($\lambda$ odd). According to Jain's composite fermion picture \cite{Jainbook}, this is the case of $2$ quasiparticles 
(2 electrons bound to $\lambda$ magnetic flux quanta each) per Landau site. The integer case $\nu=2$ has been extensively studied in the literature (see e.g. \cite{HamEzawa,EzawaBook,PRB60,PRLBrey,Schliemann,Ezawabisky,fukudamagnetotransport,zhengcanted1}), 
where the analysis of the ground state structure reveals the existence of (in general) three quantum phases, shortly denoted by: spin, canted and ppin \cite{HamEzawa,EzawaBook}, depending on which order parameter (spin or pseudospin/layer) 
dominates across the control parameter space: tunneling, Zeeman, bias, etc, couplings [see later on Eq. \eqref{HamZpZ}]. For $\nu=1$ the ground state is known to be spin-polarized in the BLQH system, that is, the canted 
phase does not exist \cite{EzawaBook}. The fractional multicomponent case (including multilayer, graphene, etc) has also been 
addressed (see e.g. \cite{zhengcanted2,PRB91,PRB75,KunYang3,nature}). In particular, for the bilayer quantum Hall (BLQH) system, the fractional case  $\nu=2/3$ has been theoretically worked out in \cite{McHal} and \cite{ezawateorico2/3}, 
having an excellent agreement with previous experimental results \cite{kumada2/3,kumada22/3,Zheng}. 

The variational states used to study the ground state and phase diagrams of (multicomponent) fractional QH systems are usually of Laughlin \cite{Laughlin} (phenomenological) type. Complementary descriptions  are provided 
by Halperin \cite{Halperin} and Haldane's \cite{Haldane} scheme of hierarchy states,  Jain's composite fermion theory \cite{Jainbook}, hierarchy states by MacDonald et al. \cite{MacDonald}, etc. 
In this article we shall use wave functions introduced previously by us in a series of papers, firstly for the bilayer case at fractional values $\nu=2/\lambda$ \cite{GrassCSBLQH,JPCMenredobicapa,JPA48} and 
recently extended to the $N$-component case at $\nu=M/\lambda$ in \cite{APsigma}. We followed a group-theoretical approach that generalizes the classification of $SU(4)$ isospin states at $\nu=2$ according to  
the 6-dimensional irreducible totally antisymmetric representation $[1,1]$ of $SU(4)$ arising in the decomposition of $4\otimes 4=10\oplus 6$. For $\nu=2/\lambda$ we constructed the $d_\lambda$-dimensional \eqref{dimension} 
irreducible representation of $SU(4)$ which, in Young tableau notation $[\lambda,\lambda]$, consists of two rows (two electrons) of $\lambda$ boxes (flux quanta) each [see eq. \eqref{CGdecomp}]. 
The corresponding phase space is the Grassmannian $\mathbb{G}^4_{2}=U(4)/[U(2)\times U(2)]$, a picture that has also been considered in some extensions to $N$-component antiferromagnets  \cite{AffleckNPB257,Sachdev,Arovas}. 
For the case of one electron, $\nu=1/\lambda$, the situation is simpler (it corresponds to a fully symmetric 
representation) and the corresponding phase space is the complex projective $\mathbb{C}P^3=U(4)/[U(3)\times U(1)]$ (the Haldane sphere $\mathbb{S}^2=\mathbb{C}P^1=U(2)/[U(1)\times U(1)]$ for the monolayer case). We shall not 
discuss the $\nu=1/\lambda$ case here. 
The dimension $d_\lambda$ of the corresponding Hilbert space at a general Landau site for $\nu=2/\lambda$ grows as $d_\lambda\sim \lambda^4$ [see eq. \eqref{dimension}] since there are more and more ways of attaching flux quanta 
to the indistinguishable electrons. These states have been used to quantify interlayer coherence and entanglement in the BLQH system at $\nu=2/\lambda$ \cite{JPCMenredobicapa}. Here we test our Grassmannian $\mathbb{G}^4_{2}$ coherent states 
as variational states to study the quantum phase diagram (according to Gilmore's algorithm \cite{Gilmore}) at filling factor $\nu=2/\lambda$, the instance $\lambda=1$ being recovered as a particular case \cite{HamEzawa,EzawaBook}. 
We shall see that the existence of the three quantum phases (spin, canted and ppin) is common to any (odd) $\lambda$, but the phase transition points depend on $\lambda$. 
We study the large isospin $\lambda\to\infty$ limit and 
realize that the canted region shrinks for high values of $\lambda$ in the tunneling direction. We also analyze the quantum case through a numerical 
diagonalization of the Hamiltonian and compare the results with the mean field (semiclassical) case. 
We obtain good agreement between quantum and semiclassical 
results in the spin and ppin phases, but not in the canted phase, where we detect exactly $\lambda$ energy-level crossings between the ground and first excited states for given values of the tunneling gap. Therefore, 
this degeneracy problem becomes more apparent for higher $\lambda$. The fidelity (overlap) between variational and numerical ground states increases when we adapt our coherent states to a parity symmetry, as turns out also to 
occur in other quantum physical systems undergoing a second order QPT like the Dicke model of atom-field interactions \cite{Dicke1,Dicke2,Dicke3}, vibron model of molecules \cite{vibron1,vibron2,vibron3,vibron4}, 
Lipkin-Meshkov-Glick \cite{Lipkin}, etc.

The organization of the paper is the following. In section \ref{sec2} we provide an oscillator realization of the $U(4)$ operators and the Landau-site Hilbert space for $\nu=2/\lambda$, together with their matrix elements, which 
will be necessary for the quantum analysis addressed in section \ref{sec4}. More details can be found in references \cite{GrassCSBLQH,JPCMenredobicapa,JPA48} and appendices \ref{basissubsec} and \ref{subsecCS}, which have 
been introduced for the sake of self-containedness. Specially appendix \ref{subsecCS}, which contains the isospin-$\lambda$ coherent states, essential for the semiclassical analysis of the model Hamiltonian studied in section \ref{sec3}. 
The Landau-site Hamiltonian governing the BLQH system at $\nu=2/\lambda$ is an adaptation of the one proposed in \cite{HamEzawa} for $\lambda=1$ to the many body case (arbitrary $\lambda$). Using our coherent states, 
we obtain in section \ref{sec3} the phase diagram (in the balanced case, for simplicity) for arbitrary $\lambda$ and, in particular, we recover the results of \cite{HamEzawa} for $\lambda=1$.
We also have a look at  $\lambda\to\infty$ (large isospin) as a formal limit. In section \ref{sec4} we analyze the quantum case through a numerical diagonalization of the Hamiltonian and compare the results with the mean field (semiclassical) case. 
For low and high interlayer tunneling (spin and ppin phases, respectively) an analytical treatment of the Hamiltonian reveals a formation of energy bands depending on angular momentum and layer population 
imbalance quantum numbers. We study the internal structure of these bands (for other studies on band structure formation in the FQHE see e.g. \cite{Jain2}).  In Section \ref{secexp} we comment on some experimental issues. Finally, 
the last section is left for conclusions and outlook.

\section{U(4) operators and Hilbert space}\label{sec2}

BLQH systems underlie an isospin $U(4)$ symmetry. In order to emphasize
the spin $SU(2)$ symmetry in the, let us say,  bottom $b$ (pseudospin down) and top $a$ or (pseudospin up)
layers, it is customary to denote the $U(4)$ generators in the four-dimensional fundamental representation by 
the sixteen $4\times 4$ matrices $\tau_{\mu\nu}\equiv\sigma_\mu^{\mathrm{ppin}}\otimes\sigma_\nu^{\mathrm{spin}}, \, \mu,\nu=0,1,2,3$, where $\sigma_\mu$ denote 
the usual Pauli matrices $\sigma_k, k=1,2,3$,  plus the identity $\sigma_0$. 
In the fractional case, bosonic magnetic flux quanta are attached to the electrons to form composite fermions. Let us denote by 
 $(a_l^\downarrow)^\dag$ [resp. $(b_l^\uparrow)^\dag$] creation operators of magnetic flux quanta (flux quanta in the sequel) attached to the electron $l$ 
with spin down [resp. up] at layer $a$ [resp. $b$], and so on. For the case of two electrons, $l=1, 2$, the four-component electron ``field'' 
$\mathcal Z$ is arranged as a compound 
$\mathcal Z=(\mathcal Z_1,\mathcal Z_2)$ of two fermions, so that the sixteen $U(4)$ density operators  are then 
written as bilinear products of creation and annihilation operators as (the so called Schwinger oscillator realization)
\begin{equation}
{T}_{\mu\nu}=\tr({\mathcal Z}^\dag\tau_{\mu\nu} \mathcal Z), \; \mathcal Z=\begin{pmatrix}
            \mathbf a\\ \mathbf b
           \end{pmatrix}=\begin{pmatrix}
\begin{matrix} a_1^\downarrow & a_2^\downarrow \\ a_1^\uparrow & a_2^\uparrow
\end{matrix}
\\ \begin{matrix} b_1^\uparrow & b_2^\uparrow\\ b_1^\downarrow & b_2^\downarrow
\end{matrix} \end{pmatrix} .\label{calzeta}\end{equation}%
 In the BLQH literature (see e.g.
\cite{EzawaBook}) it is customary to denote the total spin ${S}_k={T}_{0k}/2$ and
pseudospin ${P}_k={T}_{k0}/2$, together with the remaining 9 isospin  ${R}_{kl}={T}_{lk}/2$ operators for $k,l=1,2,3$.
A constraint in the Fock space of eight boson modes is imposed such that $\mathcal Z^\dag \mathcal Z=\lambda
I_{2}$, with $\lambda$ representing the number of magnetic flux lines piercing each electron and $I_2$ the $2\times 2$ identity. 
In particular, the linear Casimir operator ${T}_{00}=\tr(\mathcal Z^\dag \mathcal Z)$, providing the total number of flux quanta,
is fixed to $n_{a}+n_b=\lambda+\lambda=2\lambda$, with $n_{a}=n_{a1}^{\uparrow}+n_{a1}^{\downarrow}+n_{a2}^{\uparrow}+n_{a2}^{\downarrow}$ 
the total number of flux quanta in layer $a$ (resp. in layer $b$). The quadratic Casimir operator is also fixed to
\begin{equation}
\vec{S}^2 + \vec{P}^2 + \mathbf{R}^2=\lambda(\lambda+4).\label{Casimir}
\end{equation}
We also identify the interlayer imbalance operator  ${P}_3$, which measures the excess of flux quanta between
layers $a$ and $b$, that is $\frac{1}{2}(n_{a}-n_b)$. Therefore, the realization \eqref{calzeta} defines a unitary bosonic representation of the $U(4)$ matrix generators $\tau_{\mu\nu}$ in the Fock
space with constrains. This unitary irreducible representation arises  
in the Clebsch-Gordan decomposition of a tensor product of $2\lambda$ four-dimensional (fundamental, elementary) representations of
$U(4)$; for example, in Young tableau notation:
\begin{equation}
   \overbrace{\begin{Young}  \cr \end{Young}  \otimes\dots\otimes \begin{Young}  \cr \end{Young}}^{2\lambda}=
    \overbrace{\begin{Young}  & ... & \cr   & ... & \cr \end{Young}}^{\lambda}\oplus\dots, \label{CGdecomp}
\end{equation}
or $\overbrace{[1]\otimes\dots\otimes [1]}^{2\lambda}=[\lambda,\lambda]\oplus\dots$, where we wanted to highlight rectangular Young tableaux of shapes  
$[\lambda,\lambda]$ (2 rows of $\lambda$ boxes each) corresponding to $2$ electrons 
pierced by $\lambda$ magnetic flux lines (i.e., fractional filling factor $\nu=2/\lambda$). 
These are the Young tableaux determining our carrier Hilbert space ${\cal H}_\lambda(\mathbb{G}^4_2)$ 
associated to the eight-dimensional Grassmannian phase spaces $\mathbb{G}^4_{2}=U(4)/U(2)\times U(2)$ (see \cite{AffleckNPB257,Sachdev,Arovas} for similar pictures in $N$-component antiferromagnets).  
The dimension of this representation can be calculated by the hook-length formula and gives 
\begin{equation}d_\lambda=\frac{1}{12}(\lambda+1)(\lambda+2)^2(\lambda+3).\label{dimension}\end{equation}
In references \cite{GrassCSBLQH,APsigma} we have also provided a physical argument to derive the expression of $d_\lambda$ in a composite fermion picture.  
It turns out to coincide with the total number of ways to distribute $2\lambda$ flux quanta among two identical electrons in four (spin-pseudospin) states.
Note that quantum states associated to Young tableaux $[\lambda,\lambda]$ are antisymmetric (fermionic character) under the 
interchange of the two electrons (two rows) for $\lambda$ odd, whereas they are symmetric (bosonic character) for $\lambda$ even. Composite fermions require then $\lambda$ odd. 

In Refs. \cite{GrassCSBLQH,JPCMenredobicapa} we have worked out an orthonormal basis of the carrier Hilbert space
${\cal H}_\lambda(\mathbb G_2)$, which is spanned by the
set of orthonormal basis vectors
\begin{equation}
\left\{|{}{}_{q_a,q_b}^{j,m}\ra, \;\begin{matrix}
  2j, m\in\mathbb N,\\  q_a,q_b=-j,\dots,j \end{matrix}\right\}_{2j+m\leq\lambda},\label{basisvec}
\end{equation}
which can be written in terms of Fock states (to be self-contained, we give a brief in  Appendix \ref{basissubsec}). The basis states $|{}{}_{q_a,q_b}^{j,m}\ra$ turn out to be antisymmetric (resp. symmetric) 
under the interchange of the two electrons for $\lambda$ odd (resp. even), so that the parity of the number of flux quanta attached to each electron affects the quantum statistics of the compound (see \cite{JPCMenredobicapa}). 
The $d_1=6$-dimensional irrep of $SU(4)$ is usually divided into two sectors
(see e.g. \cite{EzawaBook}): the spin sector with spin-triplet pseudospin-singlet states
\begin{equation}
 |\mathfrak{S}_\uparrow\rangle=|{}{}_{\frac{-1}{2},\um}^{\,\um,\;0}\ra,\,
|\mathfrak{S}_0\rangle=\frac{1}{\sqrt{2}}(|{}{}_{\um,\um}^{\,\um, 0}\ra-|{}{}_{\frac{-1}{2},\frac{-1}{2}}^{\;\um,\;0}\ra),
\,|\mathfrak{S}_\downarrow\rangle=|{}{}_{\um,\frac{-1}{2}}^{\,\um,\;0}\ra\label{st} 
\end{equation}
and the ppin sector with pseudospin-triplet  spin-singlet  states
\begin{equation}
|\mathfrak{P}_\uparrow\rangle=|{}{}_{0,0}^{0, 1}\ra,\,
|\mathfrak{P}_0\rangle=\frac{1}{\sqrt{2}}(|{}{}_{\um,\um}^{\um,\;0}\ra+|{}{}_{\frac{-1}{2},\frac{-1}{2}}^{\;\um,\;0}\ra),
\,|\mathfrak{P}_\downarrow\rangle=|{}{}_{0,0}^{0, 0}\ra.\label{pt} 
\end{equation}

The basis states $|{}{}_{q_a,q_b}^{j,m}\ra$ are eigenstates of the following operators:
\bea
P_3|{}{}_{q_a,q_b}^{j,m}\ra&=&(2j+2m-\lambda)|{}{}_{q_a,q_b}^{j,m}\ra,\nn\\
(\vec{S}_a^2+\vec{S}_b^2)|{}{}_{q_a,q_b}^{j,m}\ra&=&2j(j+1)|{}{}_{q_a,q_b}^{j,m}\ra,\label{CCOC}\\
S_{\ell 3}|{}{}_{q_a,q_b}^{j,m}\ra&=&q_\ell|{}{}_{q_a,q_b}^{j,m}\ra,\; \ell=a, b,\nn
\eea
where we have defined angular momentum operators in layers $a$ and $b$ as  $S_{a k}=-\frac{1}{2}(S_k+R_{k3})$ and $S_{b k}=\frac{1}{2}(S_k-R_{k3})$, respectively, so that $\vec{S}_a^2+\vec{S}_b^2=\frac{1}{2}(\vec{S}^2+\vec{R}_3^2)$. 
Therefore, $j$ represents the total angular momentum of layers $a$ and $b$, whereas $q_a$ and $q_b$ are the corresponding third components. The integer $m$ is related to the interlayer imbalance (ppin third component $P_3$) through 
$\frac{1}{2}(n_{a}-n_b)=(2j+2m-\lambda)$; thus, $m=\lambda, j=0$ means $n_a=2\lambda$ (i.e., all flux quanta occupying layer $a$), whereas  $m=0, j=0$ means $n_b=2\lambda$ (i.e., all flux quanta occupying layer $b$).  
The angular momentum third components $q_a, q_b$ measure the imbalance between spin up and down in each layer, more precisely, 
$q_a=\frac{1}{2}(n_{a1}^{\uparrow}-n_{a1}^{\downarrow}+n_{a2}^{\uparrow}-n_{a2}^{\downarrow})$ and similarly for $q_b$. For later use, we shall also provide the matrix elements of the interlayer tunneling operator 
\begin{eqnarray}
&&P_1|{}{}_{q_a,q_b}^{j,m}\ra= C_{q_a,q_b}^{j,m+1}|{}{}_{q_a-\um,q_b-\um}^{j-\um,m+1}\ra+
C_{-q_a,-q_b}^{j,m+1}|{}{}_{q_a+\um,q_b+\um}^{j-\um,m+1}\ra+\nn\\ 
&&C_{-q_a+\um,-q_b+\um}^{j+\um,m+2j+2}|{}{}_{q_a-\um,q_b-\um}^{j+\um,m}\ra+
C_{q_a+\um,q_b+\um}^{j+\um,m+2j+2}|{}{}_{q_a+\um,q_b+\um}^{j+\um,m}\ra+\nn\\ 
&&C_{q_a,q_b}^{j,m+2j+1}|{}{}_{q_a-\um,q_b-\um}^{j-\um,m}\ra+
C_{-q_a+\um,-q_b+\um}^{j+\um,m}|{}{}_{q_a-\um,q_b-\um}^{j+\um,m-1}\ra+\nn\\ 
&&C_{-q_a,-q_b}^{j,m+2j+1}|{}{}_{q_a+\um,q_b+\um}^{j-\um,m}\ra+
C_{q_a+\um,q_b+\um}^{j+\um,m}|{}{}_{q_a+\um,q_b+\um}^{j+\um,m-1}\ra,\label{P1coef}
\end{eqnarray}
where the coefficients $C$ where calculated in \cite{GrassCSBLQH} and are given by
\begin{equation}
C_{q_a,q_b}^{j,m}=\frac{1}{2}\frac{\sqrt{(j+q_a)(j+q_b)m(\lambda-(m-2))}}{\sqrt{2j(2j+1)}},\; j\not=0,
\end{equation}
and $C_{q_a,q_b}^{j,m}=0$ for $j=0$.

\section{Model Hamiltonian and semiclassical analysis}\label{sec3}

Let us introduce the model Hamiltonian from first principles, in order to make clear the approximations and assumptions that we consider. More information can be found in the standard reference  \cite{EzawaBook}. 
Firstly, we consider a large cyclotron gap, so that thermal excitations across Landau levels are disregarded and electrons are confined to the lowest Landau level with vanishing kinetic energy. 
The essential properties of QH systems are determined by the Coulomb interaction Hamiltonian
\begin{equation}
 \mathcal{H}_\mathrm{C}=\frac{1}{2}\sum_{\ell,\ell'=a,b}\int d^2x d^2y V_{\ell\ell'}(x-y)\rho_\ell(x)\rho_{\ell'}(y),
\end{equation}
where $V_{\ell\ell}(r)=e^2/(4\pi\epsilon r)$ is the intralayer Coulomb interaction, whereas $V_{ab}(r)=V_{ba}(r)=e^2/(4\pi\epsilon\sqrt{r^2+\delta^2})$ is the interlayer Coulomb interaction with 
$\delta$ the interlayer separation. The Coulomb interaction is decomposed into $\mathcal{H}_\mathrm{C}=\mathcal{H}_\mathrm{C}^+ +\mathcal{H}_\mathrm{C}^-$, where $\mathcal{H}_\mathrm{C}^+$ depends on the total density $\rho=\rho_a+\rho_b$ 
while  $\mathcal{H}_\mathrm{C}^-$ depends on the density difference between layers $\Delta\rho=2\mathcal{P}_3=\rho_a-\rho_b$ (ppin third component) and is the origin of the capacitance term (see below). 
The Coulomb term $\mathcal{H}_\mathrm{C}^+$ is $SU(4)$ invariant and dominates the BLQH system provided $\delta$ is small enough (we shall usually choose $\delta=\ell_B$, the magnetic length). 

The Hamiltonian that we shall eventually use is of the sigma model type (QH ferromagnet), written in terms of collective $U(4)$ isospin operators (see \cite{APsigma} for the $N$-component case). 
Let us see how to obtain it from $\mathcal{H}_\mathrm{C}$. One proceeds by expanding the electron field operator $\psi_\mu(x)=\sum_{k}c_\mu(k)\varphi_k(x)$ in terms of one-body wave functions 
$\varphi_k(x)$ describing an electron localized around the Landau site $k$ and occupying an area of $2\pi\ell_B^2$. 
The coefficients $c_\mu(k)$ and $c_\mu^\dag(k)$ denote annihilation and creation operators of electrons with spin-ppin index $\mu=0,1,2,3$ at Landau site $k$. Substituting the expansion $\psi_\mu(x)$ into 
$\mathcal{H}_\mathrm{C}^\pm$ we obtain the Landau-site Hamiltonians
\begin{eqnarray}
 \mathcal{H}_\mathrm{C}^+&=&\sum_{klk'l'}V_{klk'l'}^+\rho(k,l)\rho(k',l'),\\
 \mathcal{H}_\mathrm{C}^-&=&4\sum_{klk'l'}V_{klk'l'}^-\mathcal{P}_3(k,l)\mathcal{P}_3(k',l'),\nonumber
\end{eqnarray}
where the Coulomb matrix elements are
\begin{equation}
 V_{klk'l'}^\pm=\frac{1}{2}\int d^2xd^2y\varphi_k^*(x)\varphi_l(x)V^\pm(x-y)\varphi_{k'}^*(y)\varphi_{l'}(y),
\end{equation}
with $V^\pm=\frac{1}{2}(V_{aa}\pm V_{ab})$, $\rho(k,l)=c^\dag(k)\tau_{00} c(l)$ is the density operator, $\mathcal{P}_3(k,l)=\frac{1}{2}c^\dag(k)\tau_{30} c(l)$ is the imbalance operator and $c=(c_0,c_1,c_2,c_3)^t$. 
In general, the $U(4)$ isospin 
operators are given by $\mathcal{T}_{\mu\nu}(k,l)=c^\dag(k)\tau_{\mu\nu} c(l)$, which is the fermionic counterpart of the bosonic representation \eqref{calzeta} for an arbitrary Landau site. 

The QH system is robust against density fluctuations; actually, we assume the suppression of charge fluctuations. Moreover, we consider the ground state $|g\ra$ to be coherent and satisfy the homogeneity 
condition $\rho(k,l)|g\ra=\nu\delta_{k,l}|g\ra$. Thus, we are working in the mean-field limit and we neglect anisotropic or translationally non-invariant solutions. Therefore, our analysis can be eventually restricted to a 
single (but arbitrary) Landau site. The direct part arising from $\mathcal{H}_\mathrm{C}^+$ is irrelevant as far as perturbations are concerned and we shall 
discard it. Therefore, we shall only consider the exchange interaction part, which can be written as a sum over isospin interactions $\delta^{\mu\mu'}\delta^{\nu\nu'}\mathcal{T}_{\mu\nu}(k)\mathcal{T}_{\mu'\nu'}(k')$ at 
Landau sites $k,k'$. Using that 
\begin{equation}
\delta^{\mu\mu'}\delta^{\nu\nu'}\mathcal{T}_{\mu\nu}\mathcal{T}_{\mu'\nu'}-\rho^2=4(\vec{\mathcal{S}}^2+\vec{\mathcal{P}}^2+
{\mathcal{R}}^2)
\end{equation}
and $\vec{\mathcal{S}}_a^2+\vec{\mathcal{S}}_b^2=\frac{1}{2}(\vec{\mathcal{S}}^2+\vec{\mathcal{R}}_3^2)$ and retaining $SU(4)$ non-invariant terms only, 
the ground state Coulomb energy per Landau site for $\nu=2$ acquires the form (when written in terms of isospin expectation values $\vec{S}$, $\vec{P}$ and $\mathbf{R}$ per Landau site) 
\begin{equation}
{H}_\mathrm{C}=4\varepsilon_{\mathrm{D}}^-P_3^2-2\varepsilon_{\mathrm{X}}^-(\vec{S}^2+\vec{R}^2_3+P_3^2),
\end{equation}
that is, a sum of the naive capacitance ($\varepsilon_{\mathrm{D}}^-$) and the exchange ($\varepsilon_{\mathrm{X}}^-$) energies. The  exchange and capacitance energy gaps are 
given in terms of the Coulomb matrix elements $V_{klk'l'}^\pm$ and, eventually, in terms of the interlayer distance $\delta$ by
\begin{equation}
\varepsilon_{\mathrm{X}}^{\pm}=\frac{1}{4}\sqrt{\frac{\pi}{2}}\left(1\pm e^{(\delta/\ell_B)^2/2}\mathrm{erfc}\left(\frac{\delta}{\sqrt{2}\ell_B}\right)\right)\mathcal{E}_C,
\end{equation}
and $\varepsilon_{\mathrm{D}}^-=\frac{\delta}{4\ell_B}\mathcal{E}_C$, where $\mathcal{E}_C=e^2/(4\pi\epsilon \ell_B)$ is the Coulomb energy unit and $\ell_B=\sqrt{\hbar c/(eB)}$ the magnetic length. 
In the following we shall simply put $\varepsilon_{\mathrm{X}}^-=\varepsilon_{\mathrm{X}}$ and $\varepsilon_{\mathrm{D}}^-=\varepsilon_{\mathrm{D}}$ as no confusion will arise. 
We shall usually choose $\delta=\ell_B$, which gives $\varepsilon_{\mathrm{X}}\simeq 0.15$ in Coulomb units.

We shall also include a (pseudo) Zeeman term
\begin{equation}
H_\mathrm{ZpZ}= -\Delta_\mathrm{Z} S_3 - \Delta_\mathrm{t} P_1 - \Delta_\mathrm{b} P_3, \label{HamZpZ}
\end{equation}
which is comprised of: Zeeman ($\Delta_\mathrm{Z}$),  interlayer tunneling  ($\Delta_\mathrm{t}$, also denoted by $\Delta_\mathrm{SAS}$ in the 
literature \cite{EzawaBook}) and bias ($\Delta_\mathrm{b}$) gaps. The bias term creates an imbalanced configuration. 
For the sake of simplicity, we shall restrict ourselves to the balanced case in the semiclassical study, which eventually means to discard the therms proportional to $\varepsilon_{\mathrm{D}}$ 
and $\Delta_\mathrm{b}$; we shall take capacitance and bias into account in the quantum analysis of section \ref{sec4}. Putting all together, the total Landau-site ground 
state energy of the BLQH system at $\nu=2$ (two electrons at a general Landau site) is \cite{HamEzawa}. 
\begin{equation}
H=H_\mathrm{C}+H_\mathrm{ZpZ}.\label{Ham}
\end{equation}

\begin{figure}[h]
\begin{center}
\includegraphics[width=\columnwidth]{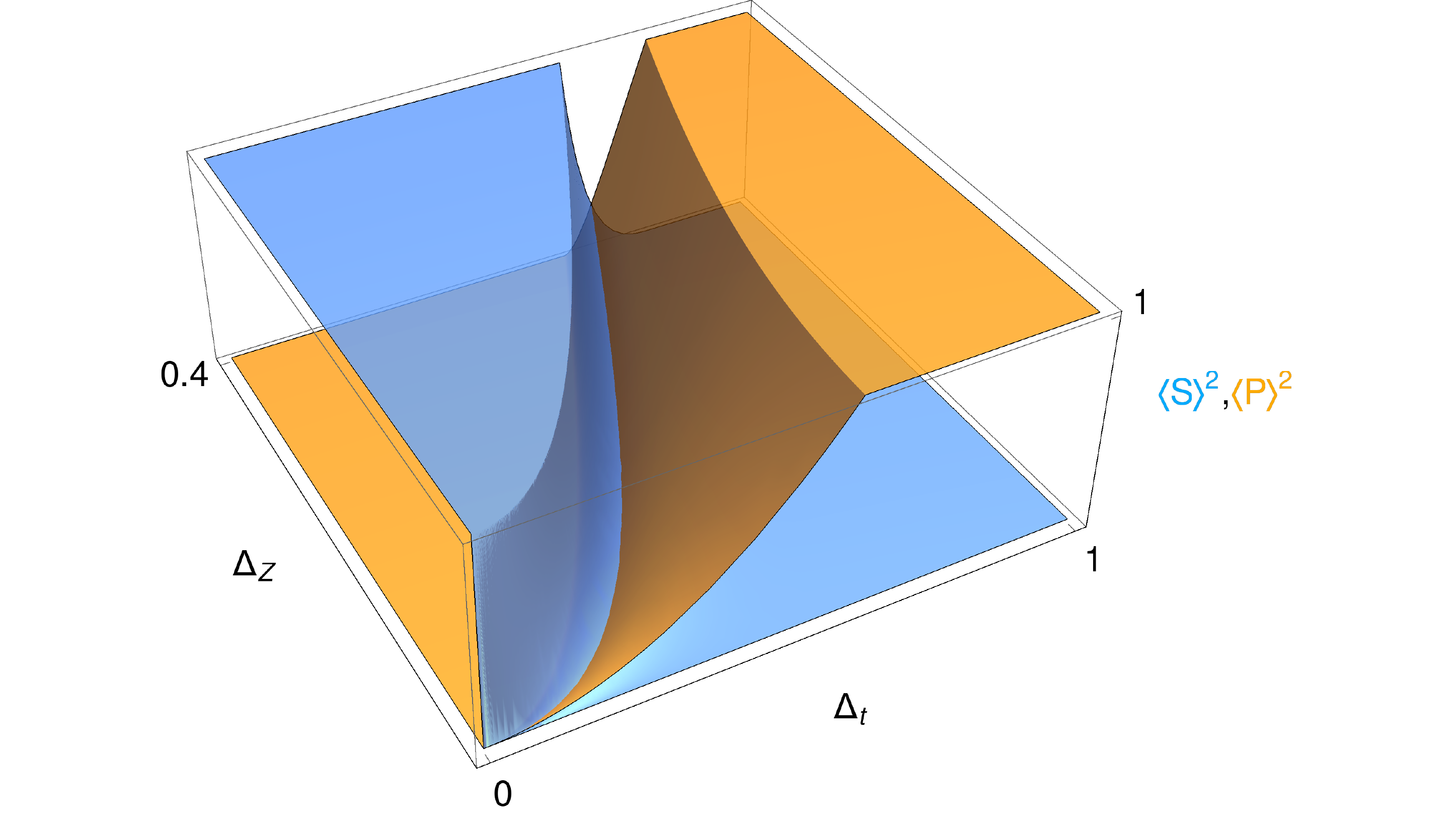}
\end{center}
\caption{Semiclassical expectation values of squared spin (blue) and ppin (orange) for $\lambda=1$ and layer separation $\delta=\ell_B$, as a function of tunneling $\DS$ and 
Zeeman $\DZ$ gaps (balanced case). We observe the three phases in table \ref{tabla}. Coulomb energy units.}\label{figfases}
\end{figure}

A minimization process of the ground state energy surface (based on a semiclassical analysis) reveals the existence of 
three quantum phases: spin, canted and pseudospin (ppin for short), which are characterized 
by the squared spin $\la\vec{S}\ra^2$ and ppin $\la\vec{P}\ra^2$ ground state mean values (order parameters), as in figure \ref{figfases} and table  \ref{tabla} for $\lambda=1$.
\begin{table} \begin{center}
 \begin{tabular}{|c|c|c|c|c|}
  \hline
 Phase:& Spin & Ppin & Canted\\ 
\hline\
Order & $\la\vec{S}\ra^2=\lambda^2$ &$\la\vec{S}\ra^2=0$ &$\la\vec{S}\ra^2\not=0$\\
parameter:& $\la\vec{P}\ra^2=0$ &$\la\vec{P}\ra^2=\lambda^2$ &$\la\vec{P}\ra^2\not=0$\\
\hline
 \end{tabular}
 \end{center}
 \caption{\label{tabla} Spin and ppin  ground state mean values in the three BLQH phases.}
\end{table}
The spin (resp. ppin) phase  occurs when the Zeeman (resp. tunneling) term dominates (see figure \ref{figfases}). The variational ground state energies of the three phases 
(spin, canted and ppin) are given by the following expressions (see \cite{HamEzawa}; we write $\Delta_{\mathrm{SAS}}=\Delta_{\mathrm{t}}$ for the sake of shortness)
\bea
E_{\mathrm{s}} & = & -2\varepsilon_{\mathrm{X}} -\Delta_{\mathrm{Z}}, \nonumber\\
E_{\mathrm{c}} & = & -2\varepsilon_{\mathrm{X}} \left(\frac{1}{4}\frac{\Delta_{\mathrm{t}}^2}{(2\varepsilon_{\mathrm{X}})^2} - \frac{1}{4}\frac{\Delta_{\mathrm{Z}}^2}{(2\varepsilon_{\mathrm{X}})^2} 
+\frac{\Delta_{\mathrm{t}}^2}{\Delta_{\mathrm{t}}^2 - \Delta_{\mathrm{Z}}^2}\right),\nonumber\\ 
E_{\mathrm{p}} & = & -\Delta_\mathrm{t},\label{enerphases}
\eea
respectively, with second order QPT critical points at $\DS^\mathrm{sc}=\sqrt{\DZ^2+4\varepsilon_{\mathrm{X}} \DZ}$ (where $E_{\mathrm{s}}=E_{\mathrm{c}}$ and $\frac{\partial E_{\mathrm{s}}}{\partial \DS}=
\frac{\partial E_{\mathrm{c}}}{\partial \DS}$) and $\DS^\mathrm{cp}=2\varepsilon_{\mathrm{X}}+\sqrt{\DZ^2+4\varepsilon_{\mathrm{X}}^2}$ (where $E_{\mathrm{c}}=E_{\mathrm{p}}$ and $\frac{\partial E_{\mathrm{c}}}{\partial \DS}=
\frac{\partial E_{\mathrm{p}}}{\partial \DS}$). 

Let us see how this phase diagram is modified for fractional filling factors $\nu=2/\lambda$. We have now $N=2\lambda$ bosonic particles (flux quanta), 
and therefore Coulomb (two-body) interactions must be renormalized by the number of pairs $N(N-1)$ to make them intensive quantities. We also divide one-body interactions by $N$ in 
order to work with energy density, as we shall have a look at the large isospin $\lambda\to\infty$ limit at some point. Taking all this information into account, we propose the following energy 
density to study the ground state at fractional filling factors $\nu=2/\lambda$:
\begin{equation}
H_\lambda=\frac{H_\mathrm{C}}{N(N-1)}+\frac{H_\mathrm{ZpZ}}{N}, \quad N=2\lambda,\label{Hamlambda}
\end{equation}
which is an adaptation of \eqref{Ham} to arbitrary $N$ flux quanta (note that $H_1=H/2$). Let us promote  $P_j, S_j$ and $R_{ij}$ to bosonic operators $T_{\mu\nu}$  in \eqref{calzeta} and consider $H_\lambda$ as an effective 
Hamiltonian per Landau site of the BLQH system at $\nu=2/\lambda$. To study the semiclassical limit, we now replace  $P_j, S_j$ and $R_{ij}$ 
by the corresponding expectation values $\la T_{\mu\nu}\ra=\la Z|T_{\mu\nu}|Z\ra$ of the operators $T_{\mu\nu}$  \eqref{calzeta} in an  
isospin-$\lambda$ coherent state $|Z\ra$ (see Appendix \ref{subsecCS}) labeled by points $Z\in\mathbb{G}^4_{2}$, i.e., $2\times 2$ complex matrices 
with four complex (eight real) entries denoted by $z^\mu=\tr(Z\sigma_\mu)/2, \mu=0,1,2,3$. Let us define ${M}_{\mu\nu}=2\ic\lambda\frac{
z_\mu \bar z^\nu-z_\nu \bar z^\mu}{\det(\sigma_0+Z^\dag Z)}$, where $z_\mu=\eta_{\mu\nu}z^\nu$  [we are using Einstein summation convention with Minkowskian
metric $\eta_{\mu\nu}=\mathrm{diag}(1,-1,-1,-1)$] and $\bar{z}^\mu$ is the complex conjugate. The coherent state expectation values of the operators appearing in the Hamiltonian \eqref{Hamlambda} 
have the following expression (see Appendix \ref{subsecCS} and references \cite{GrassCSBLQH,JPA48} for their calculation)
\bea
&&\la S_1\ra= M_{23}, \, \la S_2\ra= M_{31}, \,\la S_3\ra= M_{12},\nn\\
&&\la {R}_{k3}\ra=\ic M_{0k},\; \la \vec{S}\ra^2+\la \vec{R}_3\ra^2=M_{\mu\nu}M^{\mu\nu}/2,\label{expectv}\\
&&\la {P}_1\ra={\lambda}{\Re[\tr(Z)(1+\det(Z^\dag)]}/{\det(\sigma_0+ZZ^\dag)}, \nn\\ && \la P_3\ra= \lambda({\det(Z^\dag Z)-1})/{\det(\sigma_0+Z^\dag Z)},\nn
\eea
where $\Re$ denotes the real part [$\la {P}_2\ra$ corresponds to the imaginary part] and ``$\ic$'' is the imaginary unit. For the (cumbersome) coherent state expectation values of quadratic (two-body) operators 
$\la T^2\ra$ we address the reader to  Refs. \cite{GrassCSBLQH,JPA48}. 
Later we shall use  a parametrization \eqref{paramang} of $Z$ in terms of eight angles, for which 
the imbalance expectation value is simply $\la P_3\ra=-(\cos\vartheta_++\cos\vartheta_-)/2$. Note that the following identity for the magnitude of the $SU(4)$ 
isospin is automatically fulfilled for coherent state expectation values:
\begin{equation}
 \la \vec{S}\ra^2+\la \vec{P}\ra^2+\la \mathbf{R}\ra^2=\lambda^2.\label{isomag}
\end{equation}
For $\lambda=1$ it coincides with the variational ground state condition provided in \cite{HamEzawa}. Note the difference with the expression of the 
quadratic Casimir \eqref{Casimir}, which is fulfilled for any state of the Hilbert space. The difference between both expressions denotes the existence 
of quantum fluctuations (non-zero variance) proportional to $\lambda$ for the $SU(4)$ isospin in a coherent state. These fluctuations 
are negligible (second order) in the large isospin, $\lambda\to\infty$, classical limit. 

With these ingredients we can compute the 
energy surface $E_\lambda(Z;\varepsilon_{\mathrm{X}},\DS,\DZ)=\la Z|H_\lambda|Z\ra$ and proceed to find the values of $Z$ which minimize it. For this purpose, 
we have used the parametrization \eqref{paramang} of $Z$ in terms of eight angles (the dimension of the Grassmannian $\mathbb{G}^4_{2}$). The results of the 
minimization are as follows. For arbitrary $\lambda$, we find the same phase diagram structure as for $\lambda=1$, that is, spin, canted and ppin phases. 
In all phases we find the common relations
\begin{equation}\beta_+ = \beta_- = 0,~\vartheta_+ + \vartheta_-=\pi,~\theta_a + \theta_b = \pi,~\phi_a=\phi_b.\label{restrictangle} \end{equation}
In the spin and ppin phases we have
\begin{equation}
  \mathrm{Spin:}\; \vartheta_+^s=0=\theta_a^s,\;  \mathrm{Ppin:}\;\vartheta_+^p=-\pi/2=\theta_a^p,\label{anglespinppin}
\end{equation}
respectively. In the canted phase we get the more involved expression
\bea
\tan\vartheta_+^c &=&  \pm\sqrt{\frac{(\DS^2-\DZ^2)^2-(4\DZ \varepsilon_{\mathrm{X}}({\lambda}))^2}
{-(\DS^2-\DZ^2)^2+(4\DS \varepsilon_{\mathrm{X}}({\lambda}))^2}}, \label{thetacanted}\\
\tan\theta_b^c &=& \mp\frac{\DS}{\DZ} \sqrt{\frac{(\DS^2-\DZ^2)^2-(4\DZ \varepsilon_{\mathrm{X}}({\lambda}))^2}
{-(\DS^2-\DZ^2)^2+(4\DS \varepsilon_{\mathrm{X}}({\lambda}))^2}},\nn
\eea
where we have defined $\varepsilon_{\mathrm{X}}({\lambda})=\lambda\varepsilon_{\mathrm{X}}/(2\lambda-1)$ for later use. Note that we have two different solutions of $(\vartheta_+^c,\theta_b^c)$ in the canted phase, given by the 
signs $(+,-)$ and $(-,+)$ in equation \eqref{thetacanted}, leading to the same minimum energy $\la Z^c_\pm|H_\lambda|Z^c_\pm\ra$, with $Z^c_\pm=Z(\theta_{a,b},\phi_{a,b},\vartheta_\pm,\beta_\pm)|_\pm^c$ 
the corresponding stationary point in the Grassmannian $\mathbb{G}^4_{2}$ for any of the two solutions  $(+)=(+,-)$ and $(-)=(-,+)$ together with the common restrictions \eqref{restrictangle}. 
Even though both coherent states $|Z^c_+\ra$ and $|Z^c_-\ra$ give the same energy, they are 
distinct; in fact, they are almost orthogonal $\la Z^c_+|Z^c_-\ra\simeq 0$ in the canted phase. This indicates that the ground state is degenerated and there is a broken symmetry in the thermodynamic limit. We will come back 
to these degeneracy problems of the canted phase in the next section.

\begin{figure}[h]
\begin{center}
\includegraphics[width=\columnwidth]{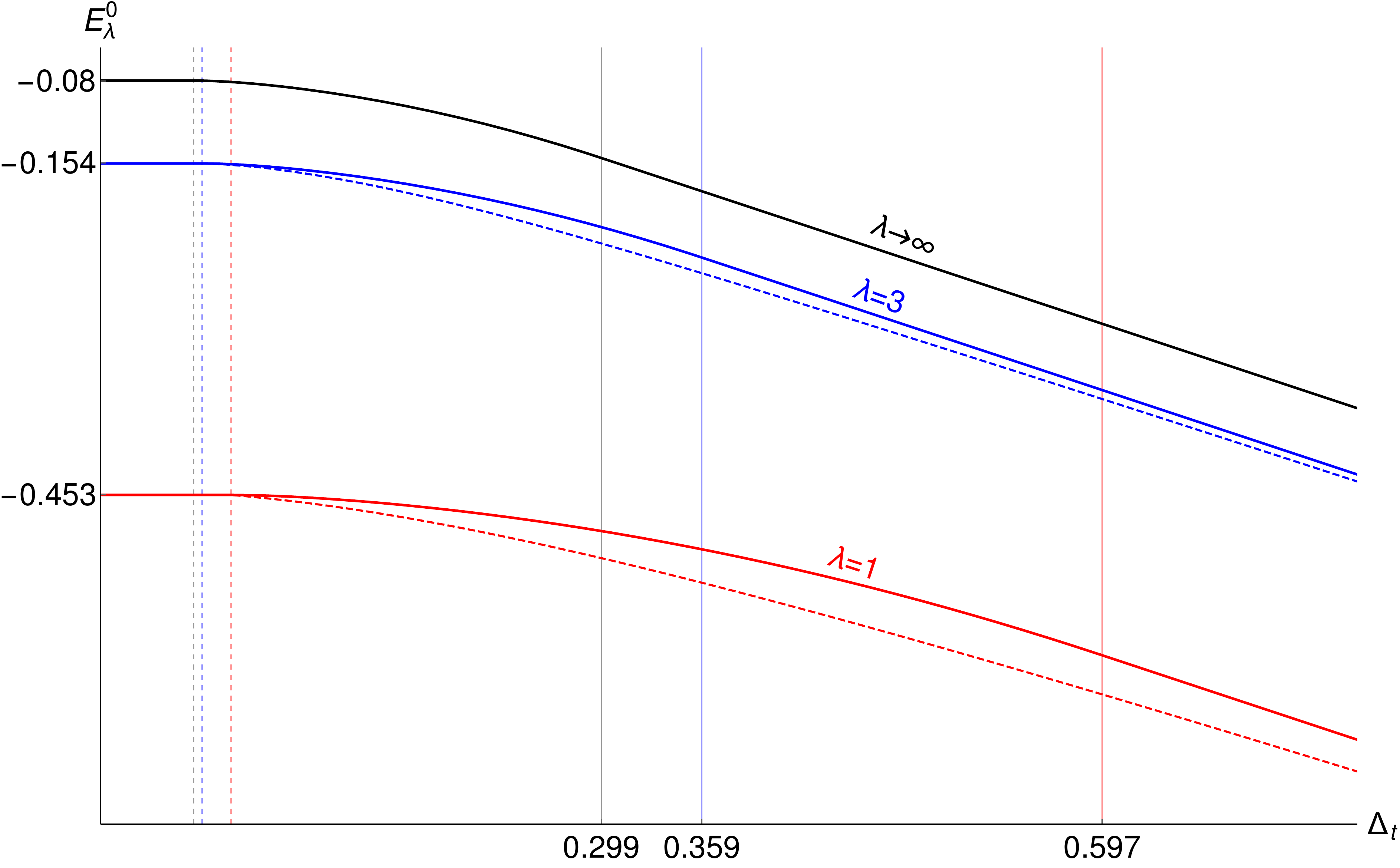}
\end{center}
\caption{Variational (solid) and exact (dotted) ground state energy $E_\lambda^0$ as a function of tunneling $\DS$ (fixed layer separation $\delta=\ell_B$ and Zeeman $\DZ=0.01$) for $\lambda=1$ (red), 
$\lambda=3$ (blue) and  $\lambda\to\infty$ (dotted black). 
The corresponding spin-canted $\Delta^{\mathrm{sc}}_\mathrm{t}(\lambda)$ and canted-ppin $\Delta^{\mathrm{cp}}_\mathrm{t}(\lambda)$ 
phase transition points are represented by vertical dashed and solid grid lines, respectively. Numerical values for the canted-ppin transition points $\Delta^{\mathrm{cp}}_\mathrm{t}(\lambda)$ are also indicated. For the 
spin-canted transition we have $\Delta^{\mathrm{sc}}_\mathrm{t}(1)=0.078$, $\Delta^{\mathrm{sc}}_\mathrm{t}(3)=0.061$ and $\Delta^{\mathrm{sc}}_\mathrm{t}(\infty)=0.056$. Coulomb energy units.}\label{figenergy}
\end{figure}

Let us denote collectively by $Z^0_+$ and $Z^0_-$ the two sets of stationary points in any of the three (spin, canted and ppin) quantum phases (note that $Z^0_+=Z^0_-$ in the spin and ppin phases). 
Both sets of stationary points provide the same value of the energy expectation value $E_\lambda(Z^0_\pm;\varepsilon_{\mathrm{X}},\DS,\DZ)=\la Z^0_\pm|H_\lambda|Z^0_\pm\ra$, which we shall simply denote  
by $E_\lambda^0(\varepsilon_{\mathrm{X}},\DS,\DZ)$. After some algebraic manipulations, one can see that $E_\lambda^0$  coincides with \eqref{enerphases} when 
replacing $\varepsilon_{\mathrm{X}}\to \varepsilon_{\mathrm{X}}({\lambda})$, except for a zero-point energy correction $\mathcal{E}^0_\lambda=-2\varepsilon_{\mathrm{X}}/(2\lambda-1)$. This zero-point energy is just due to the non-zero 
quantum fluctuations $\la A^2\ra\not=\la A\ra^2$ of $SU(4)$ operators [compare for example \eqref{Casimir} with \eqref{isomag}] and it vanishes in the high isospin $\lambda$ limit. 
There is also a normalization factor of two difference since, for $\lambda=1$, $H_\lambda$ in \eqref{Hamlambda} is related to $H$ in 
\eqref{Ham} by $H_1=H/2$. In Figure \ref{figenergy} we represent the variational energy density $E_\lambda^0$ as a function of 
$\DS$ for $\DZ=0.01$ and interlayer distance $\delta=\ell_B$ for different values of $\lambda$. We see that the spin, canted and ppin phase regions (separated by vertical grid lines) are 
affected by the value of the isospin $\lambda$; in fact,  the new critical points are displaced at 
\begin{eqnarray}
 \DS^\mathrm{sc}(\lambda)&=&\sqrt{\DZ^2+4\varepsilon_{\mathrm{X}}(\lambda) \DZ},\;\label{critpnt}\\ 
\DS^\mathrm{cp}(\lambda)&=&2\varepsilon_{\mathrm{X}}(\lambda)+\sqrt{\DZ^2+4\varepsilon_{\mathrm{X}}^2(\lambda)},\nn
\end{eqnarray}
coinciding with the ones after eq. \eqref{enerphases} when replacing $\varepsilon_{\mathrm{X}}\to \varepsilon_{\mathrm{X}}({\lambda})$.

 The high isospin limit, $\lambda\to\infty$ is also formally and 
straightforwardly accomplished just by replacing $\varepsilon_{\mathrm{X}}\to \varepsilon_{\mathrm{X}}(\infty)= \varepsilon_{\mathrm{X}}/2$ in the energy \eqref{enerphases} and critical points 
$\DS^\mathrm{sc}$ and $\DS^\mathrm{cp}$. We see that the width of the canted region $\DS^\mathrm{cp}(\lambda)-\DS^\mathrm{sc}(\lambda)$ shrinks as $\lambda$ increases. We also compare in figure \ref{figenergy} 
the variational (solid) and numerical (dashed) ground state energies (see next section for a quantum analysis). We realize that the variational and numerical results coincide in the spin phase but not in the 
canted and ppin regions, except in the high isospin $\lambda$ limit, where exact results converge to the semiclassical (mean-field) limit. We must stress that the large $\lambda$ limit is considered here as a 
formal, mathematical, reference limit only. To consider large $\lambda$ as a physical limit, we should relax some of the assumptions and approximations made to arrive to the model Hamiltonian $H_\lambda$ concerning, for example, 
the charge gap.

\begin{figure}[h]
\begin{center}
\includegraphics[width=\columnwidth]{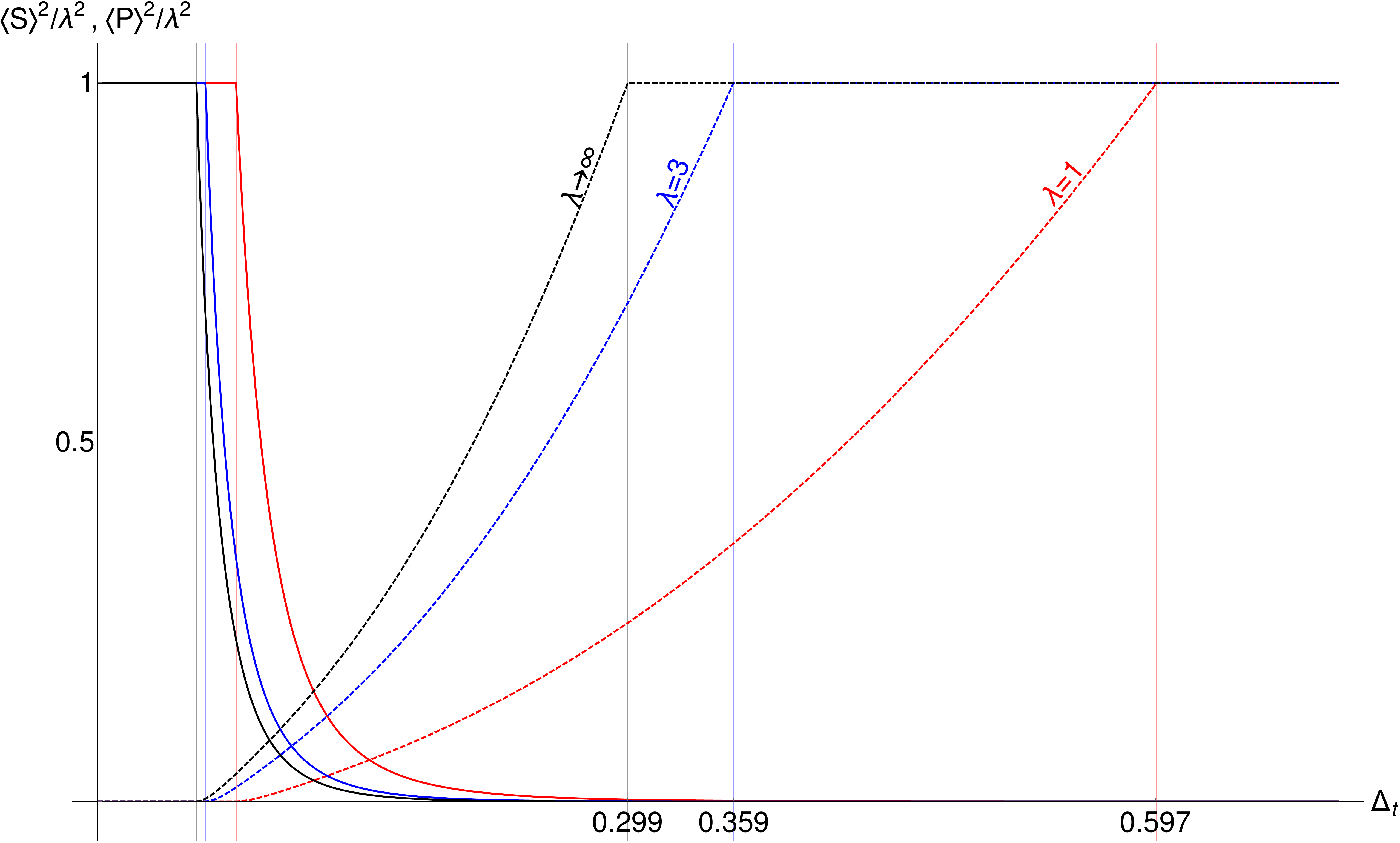}
\end{center}
\caption{Semiclassical expectation values of squared spin (solid) and ppin (dashed) for $\lambda=1$ (red), $\lambda=3$ (blue) and $\lambda\to\infty$ (black) for layer separation $\delta=\ell_B$ and $\DZ=0.01$, 
as a function of tunneling $\DS$. Vertical grid lines indicate the spin-canted and canted-ppin (with numerical value) phase transition points for each $\lambda$. Coulomb energy units.}\label{figspin}
\end{figure}

Spin-canted and canted-ppin phase transition points are better appreciated in Figure \ref{figspin}, where we represent normalized squared spin $\la \vec{S}\ra^2$ and ppin $\la \vec{P}\ra^2$ order parameters for the variational 
coherent states $|Z^0_\pm\ra$  as a function of 
$\DS$ for $\DZ=0.01$, $\delta=\ell_B$ and different values of $\lambda$. An explicit expression of  spin $\la \vec{S}\ra^2$ and ppin $\la \vec{P}\ra^2$ can be easily 
obtained from the expectation values \eqref{expectv}, together with the restrictions \eqref{restrictangle} common to the three phases, resulting in 
\bea
\la\vec{S}\ra^2&=&\lambda^2\cos^2\vartheta_+\cos^2\theta_b,\nn\\
\la\vec{P}\ra^2&=&\lambda^2\sin^2\vartheta_+\sin^2\theta_b. \label{spinppin}\nn
\eea
In the spin phase ($\vartheta_+=0=\theta_b$) we have maximum spin $\la \vec{S}\ra^2=\lambda^2$ [remember the identity \eqref{isomag}] and minimum ppin $\la \vec{P}\ra^2=0$, whereas 
in the ppin phase ($\vartheta_+=-\pi/2=\theta_b$) we have minimum spin $\la \vec{S}\ra^2=0$ and maximum ppin $\la \vec{P}\ra^2=\lambda^2$. In the canted phase, when inserting 
\eqref{thetacanted} into \eqref{spinppin}, we realize that both spin and ppin do not 
attain the maximum value, as can be appreciated in figure \ref{figspin} and summarized in table \ref{tabla} [the case $\lambda=1$ was already depicted in figure \ref{figfases}]. 
Note that both (negative and positive) values of $\theta_a^c$ and $\vartheta_+^c$ in \eqref{thetacanted} 
give the same values of energy and squared spin and ppin in \eqref{spinppin}, even though the corresponding variational states $|Z^c_-\ra$ and $|Z^c_+\ra$ are different (quasi-orthogonal). This reflects 
a degeneracy problem that we shall analyze in the following section.

\section{Quantum analysis and numerical diagonalization results}\label{sec4}

In this section we solve the eigenvalue problem for the Hamiltonian \eqref{Hamlambda} and compare with the mean field (semiclassical) results of the previous section, analyzing the effect 
of quantum fluctuations. 
The Hamiltonian matrix elements in the basis \eqref{basisvec} are determined by the expressions \eqref{CCOC} and \eqref{P1coef}. 
For example, for $\lambda=1$ and arranging the basis vectors \eqref{basisvec} as 
\bea
| 1 \ra = |{}{}_{0,0}^{0,0}\ra,\;| 2 \ra = |{}{}_{0,0}^{0,1}\ra,\;| 3 \ra = |{}{}^{\frac{1}{2},~~0}_{\frac{-1}{2},\frac{-1}{2}}\ra,\nn\\
| 4 \ra = |{}{}^{\frac{1}{2},0}_{\frac{1}{2},\frac{1}{2}}\ra,\;| 5 \ra = |{}{}^{\frac{1}{2},~0}_{\frac{1}{2},\frac{-1}{2}}\ra,\; | 6 \ra = |{}{}^{\frac{1}{2},~0}_{\frac{-1}{2},\frac{1}{2}}\ra,
\eea
we obtain the $6\times 6$ Hamiltonian matrix
\begin{widetext}
\begin{equation}
H_1=\left(
\begin{array}{cccccc}
 \frac{1}{2} (\Delta_\mathrm{b}+4 \varepsilon_\mathrm{D}-2 \varepsilon_\mathrm{X}) & 0 & -\frac{\DS }{4} & -\frac{\DS }{4} & 0 & 0 \\
 0 & \frac{1}{2} (-\Delta_\mathrm{b}+4 \varepsilon_\mathrm{D}-2 \varepsilon_\mathrm{X})  & -\frac{\DS }{4} & -\frac{\DS }{4} & 0 & 0 \\
 -\frac{\DS }{4} & -\frac{\DS }{4} & -3 \varepsilon_\mathrm{X} & 0 & 0 & 0 \\
 -\frac{\DS }{4} & -\frac{\DS }{4} & 0 & -3 \varepsilon_\mathrm{X} & 0 & 0 \\
 0 & 0 & 0 & 0 & - \frac{1}{2} (\DZ+6 \varepsilon_\mathrm{X}) & 0 \\
0 & 0 & 0 & 0 & 0 &  \frac{1}{2} (\DZ -6 \varepsilon_\mathrm{X}).\\
\end{array}
\right)\end{equation}
\end{widetext}
Those readers more acquainted with the spin-triplet (ppin-singlet) and ppin-triplet (spin-singlet) states can perform the change of basis \eqref{st} and \eqref{pt}. 
The lowest (ground state) energy $E^0_\lambda$ is plotted in figure \ref{figenergy} (dotted curves) as a function of $\Delta_\mathrm{t}$ for $\lambda=1$ and $\lambda=3$ 
(Hilbert space dimensions $d_1=6$ and $d_3=50$, respectively). As we have already commented, the exact ground state energy coincides with the mean-field result in the spin phase. Actually, the lowest energy eigenstate 
in the spin phase is the basis state $|{}{}_{q_a,q_b}^{j,m}\ra=|{}{}_{-\lambda/2,\lambda/2}^{\lambda/2,\quad 0}\ra$, which is also a extremal case of coherent state $|Z\ra$ for the critical angle values 
\eqref{restrictangle} and \eqref{anglespinppin} in the spin phase. The mean-field result does not coincide with the numerical diagonalization in the  canted and ppin phases, but the energy difference between both 
gets smaller as $\lambda$ increases, as can be appreciated in figure \ref{figenergy}.

\begin{figure}[h]
\begin{center}
\includegraphics[width=\columnwidth]{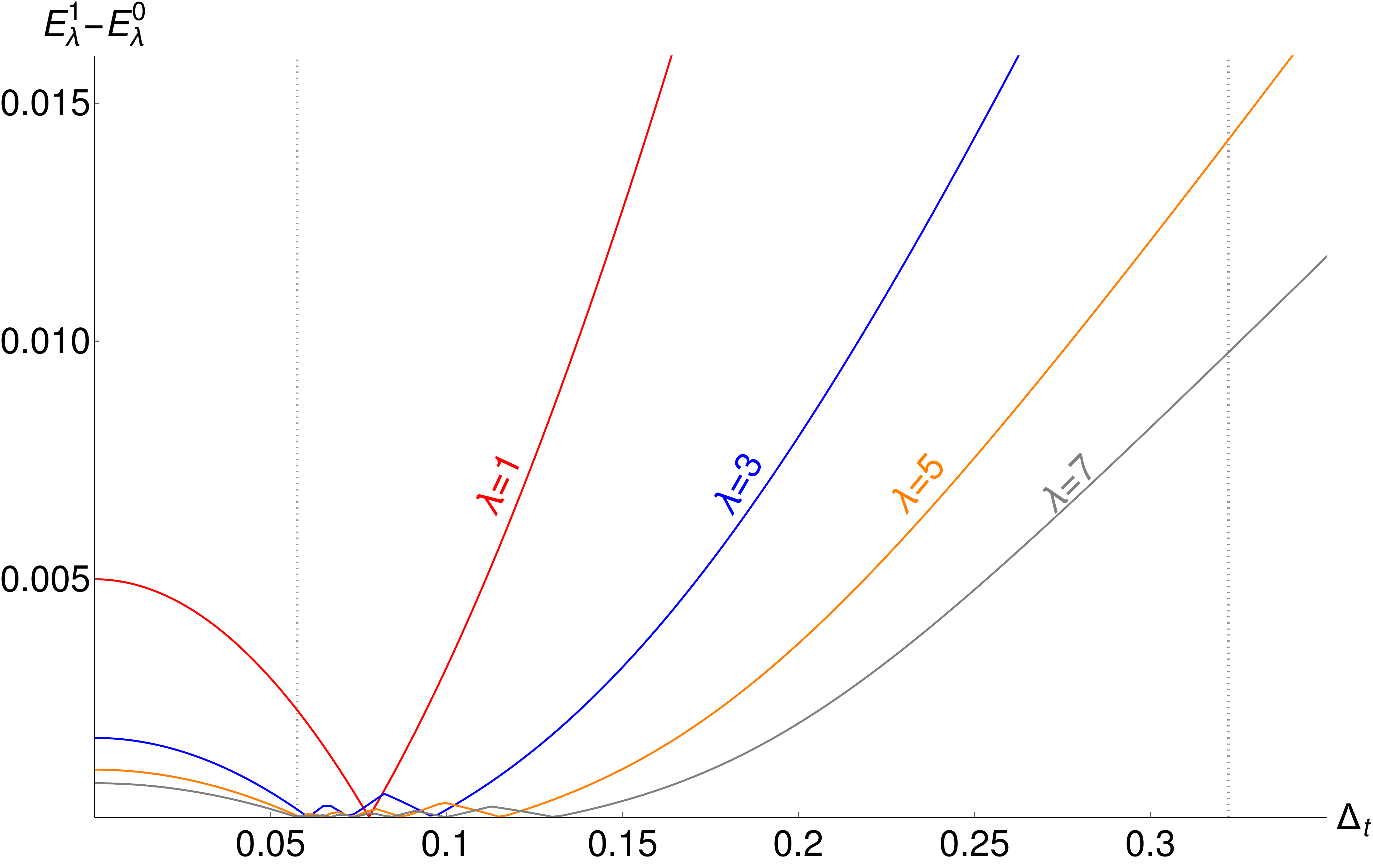}
\end{center}
\caption{Energy gap between ground and first energy levels as a function of $\DS$, for $\delta=\ell_B$, $\DZ=0.01$ and different values of $\lambda$. As a reference, we include vertical grid lines indicating  
the spin-canted $\DS^\mathrm{sc}(\infty)=0.0556$ and canted-ppin $\DS^\mathrm{cp}(\infty)=0.299$ transition points for the limiting case $\lambda\to\infty$. Coulomb energy units.}\label{figdeg}
\end{figure}

\begin{figure}[h]
\begin{center}
\includegraphics[width=\columnwidth]{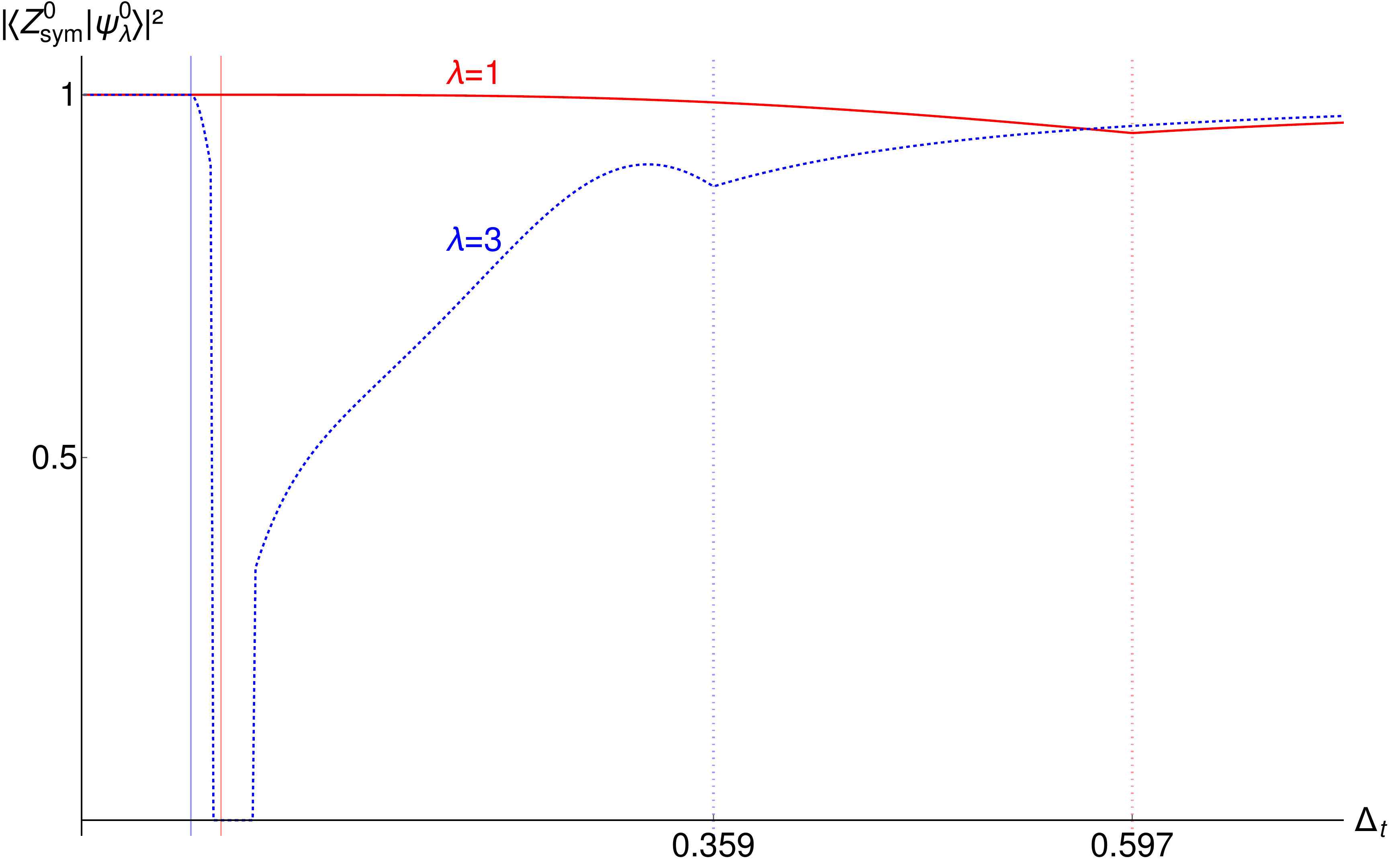}
\end{center}
\caption{Overlap/fidelity $|\la Z^0_\mathrm{sym}|\psi^0_\lambda\ra|^2$ between variational (coherent states) adapted to the parity symmetry and exact ground states, as a function of $\DS$, for $\delta=\ell_B$, $\DZ=0.01$, 
$\lambda=1$ (red) and $\lambda=3$ (blue-dashed). Spin-canted and canted-ppin (with numerical value) phase transition points are indicated by vertical solid and dotted grid lines, respectively. Coulomb energy units. }\label{figfid}
\end{figure}

In figure \ref{figspinq} we represent the exact (red) variational (dotted-black) and parity-adapted (blue; see below)  ground state squared expectation values of spin $\la \vec{S}\ra^2$  (solid) and ppin $\la \vec{P}\ra^2$  (dashed) 
as a function of $\DS$ for $\lambda=3$. The variational case was already depicted  in figure \ref{figspin} and presents a smooth behavior, which contracts with the step-like behavior in the quantum case, mainly 
in the canted phase, where we find in general $\lambda$ steps for $\la \vec{S}\ra^2$ and $\la \vec{P}\ra^2$. Moreover, the transition from canted to ppin phase is not so well marked as the transition from spin to canted phase, 
which occurs quite sharply. This result agrees with the one obtained in \cite{Schliemann} through an exact diagonalization of a few-electron system, where the boundary between the spin and canted phases
is practically unmodified from the mean-field result, but the boundary between the
canted and ppin phases is considerably modified. 
\begin{figure}[h]
\begin{center}
\includegraphics[width=\columnwidth]{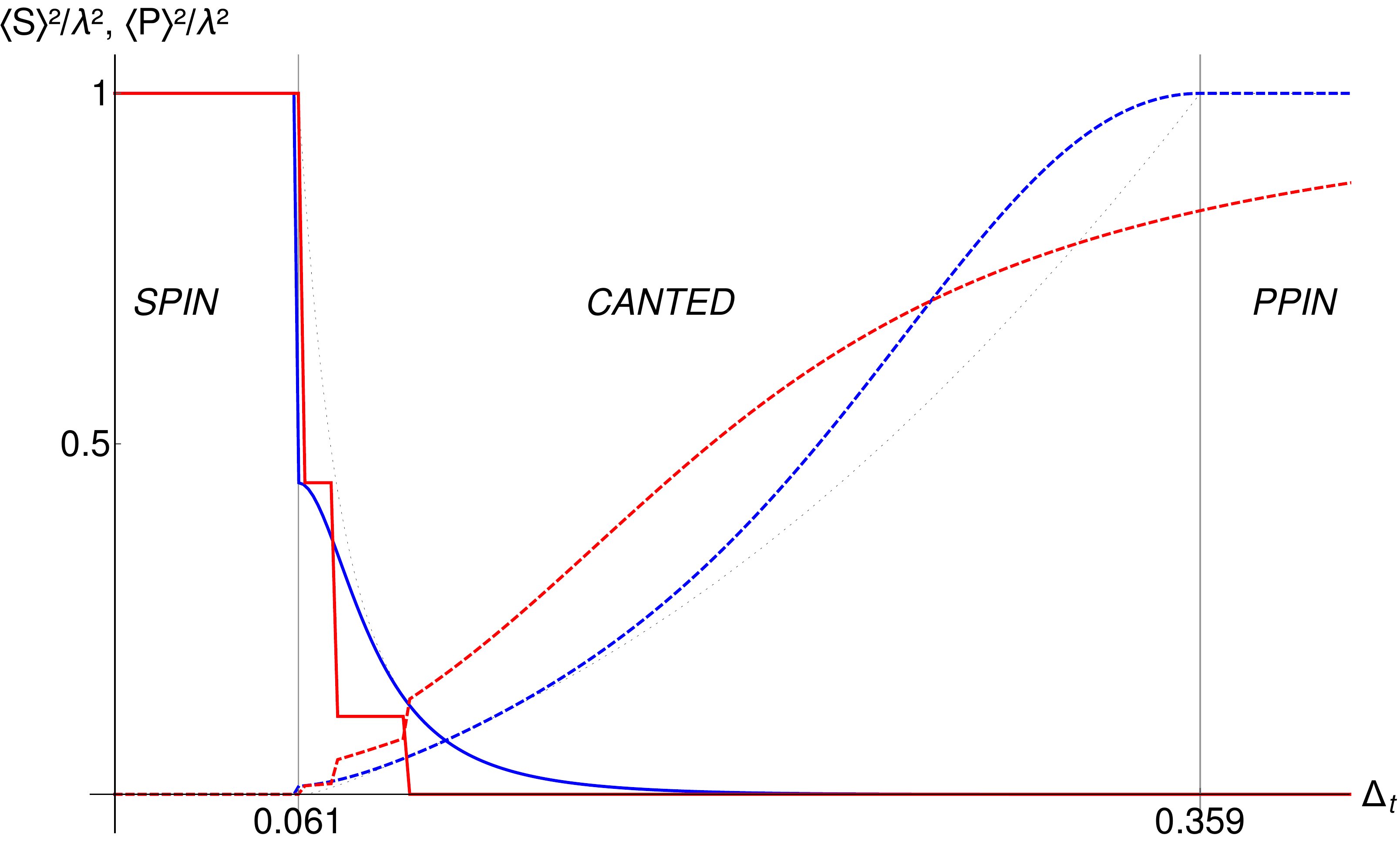}
\end{center}
\caption{Exact (red) variational (dotted-black) and parity-adapted (blue)  ground state squared expectation values of spin $\la \vec{S}\ra^2$  (solid) and ppin $\la \vec{P}\ra^2$  (dashed) 
for $\lambda=3$, layer separation $\delta=\ell_B$ and $\DZ=0.01$, 
as a function of tunneling $\DS$. Vertical grid lines indicate the spin-canted $\DS^\mathrm{sc}(3)=0.061$ and canted-ppin $\DS^\mathrm{cp}(3)=0.359$ phase transition points for $\lambda=3$. Coulomb energy units.}\label{figspinq}
\end{figure}
The step-like behavior of spin and ppin in the canted phase is due to a level crossing at certain values of the tunneling $\DS$ for which the ground and first excited energy levels degenerate. The number of 
level crossings increases with $\lambda$, in fact, there are exactly $\lambda$ crossings (see figure \ref{figdeg}). In the high isospin $\lambda$ limit, this might indicate that there is an 
avoided crossing in the whole canted region.
This degenerate situation makes that the overlap between the variational (mean-field)  $|Z^0_\pm\ra$ and exact (numerical) $|\psi^0_\lambda\ra$ 
ground states is quite small and irregular in the canted phase. We get better results (but still not good enough) by adapting our variational states to the parity symmetry, that is, by taking the 
normalized symmetric combination
\begin{equation}
 |Z^0_\mathrm{sym}\ra=\frac{|Z^0_+\ra+|Z^0_-\ra}{\sqrt{2(1+\Re(\la Z^0_+|Z^0_-\ra})}.\label{parityadapt}
\end{equation}
The results of the overlap/fidelity $|\la Z^0_\mathrm{sym}|\psi^0_\lambda\ra|^2$ between variational and exact ground states is shown in figure \ref{figfid}. We see that 
the fidelity is 1 in the spin phase (where the variational and exact ground states coincide with $|{}{}_{-\lambda/2,\lambda/2}^{\lambda/2,\quad 0}\ra$) 
and less that 1 in the ppin phase (although it increases with $\DS$). The degenerate situation in the canted phase gives low fidelity, except for $\lambda=1$. 
A fidelity drop is also expected at the phase transition point, where quantum fluctuations dominate. For completeness, we have also represented in figure \ref{figspinq} the 
squared expectation values of spin $\la Z^0_\mathrm{sym}|\vec{S}|Z^0_\mathrm{sym}\ra^2$ (solid-blue curve) and ppin $\la Z^0_\mathrm{sym}|\vec{P}|Z^0_\mathrm{sym}\ra^2$ (dashed-blue) in the 
parity-adapted state \eqref{parityadapt}, which `interpolates' between the semiclassical and the quantum case.

Parity adapted coherent states like \eqref{parityadapt} have also been successfully used to better reproduce the exact quantum results at finite-size from the mean-field approximation in 
other interesting models undergoing a second order QPT like for example the Dicke model of atom-field interactions \cite{Dicke1,Dicke2,Dicke3}, the vibron model of molecules \cite{vibron1,vibron2,vibron3,vibron4} 
and the Lipkin-Meshkov-Glick model \cite{Lipkin}.

\begin{figure}[h]
\begin{center}
\includegraphics[width=\columnwidth]{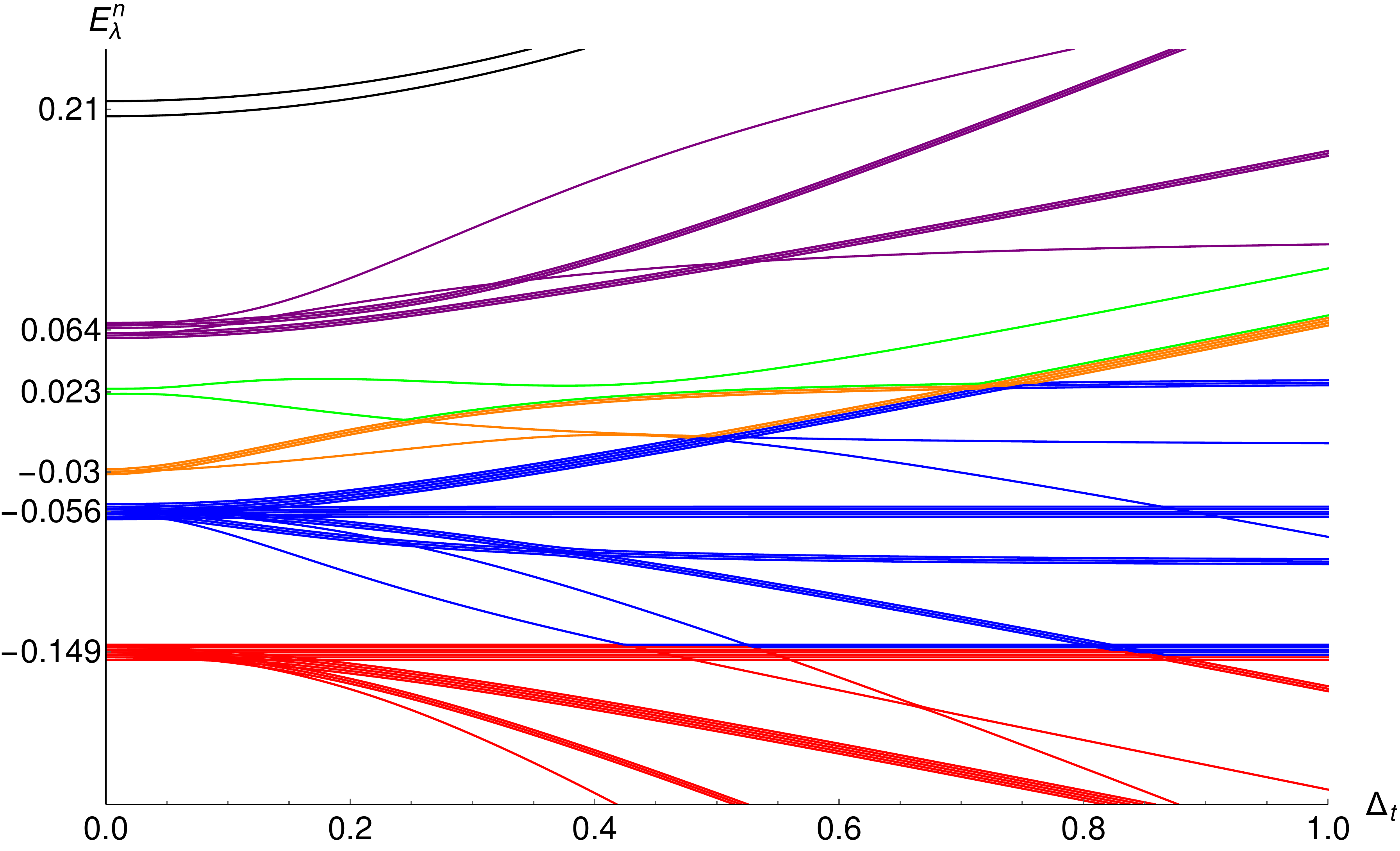}
\end{center}
\caption{Energy spectrum as a function of $\DS$, for $\delta=\ell_B$, $\DZ=0.01$, $\Delta_\mathrm{b}=0.01$ and $\lambda=3$. The six energy bands emerge in the spin phase (small $\DS$). They are plotted in 
different colors and its structure is given in table \ref{tablenergy}. Coulomb energy units.}\label{figbandas}
\end{figure}

So far we have only studied the balanced case and restricted our analysis to the ground (and first excited) state. For completeness, let us introduce imbalance (capacitive and bias terms) and have a look to 
the whole spectrum. In figure \ref{figbandas} we represent the energy spectrum as a function of $\DS$ for $\lambda=3$ (dimension $d_3=50$). 

For zero tunneling $\DS=0$ (spin phase) we observe a band energy structure that we can explain 
as follows. The Hamiltonian for this case is diagonal in the basis \eqref{basisvec} and the corresponding eigenvalues can be straightforwardly obtained from \eqref{CCOC} as
\begin{eqnarray}
E_\lambda({}{}^{j,m}_{q_a,q_b})&=&\frac{\varepsilon_\mathrm{cap}(2j+2m-\lambda)^2-8\varepsilon_\mathrm{X}j(j+1)}{2\lambda(2\lambda-1)}\nn\\ 
&&-\frac{\DZ(q_b-q_a)+\Delta_\mathrm{b}(2j+2m-\lambda)}{2\lambda},
\end{eqnarray}
where $\varepsilon_\mathrm{cap}=4\varepsilon_\mathrm{D}-2\varepsilon_\mathrm{X}$ denotes the capacitance energy. For small Zeeman and bias interactions, the dominant part are the capacitance and 
exchange energies which grow with the squared angular momentum $j(j+1)$ and the squared layer population imbalance $\varsigma^2=(2j+2m-\lambda)^2$. These two magnitudes $(j,\varsigma)$ 
roughly determine (with some exceptions for higher $\lambda$) the energy band arrangement at zero tunneling. In table \ref{tablenergy} we represent the band energy structure, giving a representative value of the band energy 
$E^{j,\varsigma}_\lambda$ together with the values of the angular momentum $j$ and absolute imbalance  $\varsigma=|2j+2m-\lambda|$ common to each energy band for $\lambda=3$. 
In general, there are two values of $m_\pm=(\pm\varsigma+\lambda-2j)/2$ and $(2j+1)^2$ values of $q_a,q_b$ common to every couple $(j,\varsigma)$, so that 
the total number of levels forming the energy band $(j,\varsigma)$ is $2(2j+1)^2$, except for $\varsigma=0$ that is just $(2j+1)^2$. 
Small bias and Zeeman interactions slightly break the degeneracy in $m$ and $q_a,q_b$, respectively, determining the bandwidth. 

\begin{table} \begin{center}
 \begin{tabular}{|c|c|c|c|}
  \hline
 $E^{j,\varsigma}_\lambda$ & $j,\varsigma$ & No. levels \\ 
\hline\
-0.15 & $j=3/2, \varsigma=0$ & 16\\
-0.056& $j=1, \varsigma=1$ & 18 \\
-0.03& $j=1/2, \varsigma=0$ & 4\\
0.023& $j=0, \varsigma=1$ & 2\\
0.064& $j=1/2, \varsigma=2$ & 8\\
0.21& $j=0, \varsigma=3$ & 2\\
\hline\
& Total:& 50\\
\hline
 \end{tabular}
 \end{center}
 \caption{\label{tablenergy} Representative value of the band energy $E^{j,\varsigma}_\lambda$ at $\DS=0$ (spin phase) for $\lambda=3$ and $\DZ=0.01=\Delta_\mathrm{b}$. Each energy band $(j,\varsigma)$ 
 is roughly determined by angular momentum and layer population imbalance $\varsigma=|2j+2m-\lambda|$. 
 There are $2(2j+1)^2$ closely spaced energy levels  forming each energy band $(j,\varsigma)$, except for $\varsigma=0$, where there are only $(2j+1)^2$. Coulomb energy units.}
\end{table}

For large tunneling $\DS$ (ppin phase), energy bands are formed around the eigenvalues of the ppin first component $P_1$, that is, at energies about $E_\lambda^{j+m}=-\DS (2j+2m-\lambda)/(2\lambda)$, since the eigenvalues of 
$P_1$ and $P_3$ coincide (actually, the discussion also applies for large bias voltage). Therefore, the number of energy bands arising at high $\DS$ is exactly $2\lambda+1$ 
[remember discussion in paragraph between \eqref{CCOC} and \eqref{P1coef}]. We can label each band by $n=2j+2m$ [the homogeneity degree of polynomials \eqref{basisfunc}] and 
the number of closely spaced energy levels forming each energy band $n$ is
\begin{equation}
D_n=\left\{\ba{l} \frac{(n+1)(n+2)(n+3)}{6}, \,n\leq\lambda,\\ \frac{(2\lambda-n+1)(2\lambda-n+2)(2\lambda-n+3)}{6},\, 
\lambda\leq n\leq 2\lambda.\ea\right.
\end{equation}
One can verify that $\sum_{n=0}^{2\lambda}D_n={d_\lambda}$ gives the dimension of the Hilbert space. At intermediate tunneling (canted phase) there is an intricate spectrum structure with multiple band-crossing.

\section{Comparison with experiments}\label{secexp}

In this article we are using a quite simplified (toy) model, with the assumptions and approximations stated at the beginning of section \ref{sec3}. Our main aim is to promote the Hilbert space of composite fermions at one Landau 
site for $\nu=2/\lambda$, focusing on the structure of the quantum phases and their boundaries. Therefore, here we just aspire to capture the essence and to give a qualitative description of some experimental data and 
basic phenomenology. A more faithful description of real BLQH systems would require a more sophisticated model taking into account interactions.

That said, let us comment about some experimental issues in connection with Kumada's et al. results in references  \cite{kumada2/3} and \cite{kumada22/3}, about BLQH systems (ussually GaAs/
AlGaAs double-quantum-well samples)  at $\nu=2$ and $\nu=2/3$, respectively, and 
references \cite{ezawateorico2/3} and \cite{Zheng} for $\nu=2/3$. In  \cite{kumada2/3,kumada22/3} the authors 
provide a relation between Zeemann $\Delta_\mathrm{Z}$ and tunneling $\Delta_\mathrm{t}$ gaps in the spin-ppin phase transition. They observe that $\Delta_\mathrm{Z}$ must be enhanced by a factor of 10 with respect to 
$\Delta_\mathrm{t}$  for $\nu=2/3$ and by a factor of 20 for $\nu=2$. In other words, they observe that, for a fixed $\Delta_\mathrm{Z}$, the phase transition for $\nu=2/3$ occurs for 
lower values of $\Delta_\mathrm{t}$ than for $\nu=2$. This behavior is qualitatively captured by our results in equation \eqref{critpnt} and figures \ref{figenergy} and \ref{figspin}, which show that 
the critical point  $\DS^\mathrm{cp}(\lambda)$ decreases with $\lambda$ for fixed $\Delta_\mathrm{Z}$ (the canted region shrinks). 
This enhancement of $\Delta_\mathrm{Z}$ and suppression of $\Delta_\mathrm{Z}$ is claimed to be due to interaction effects between composite fermions and between electrons. 
Therefore, a better fit could perhaps be obtained with a less simplified model. 

The analysis of the energy difference between the two lowest eigenstates of a BLQH system at $\nu=2/3$ made in reference \cite{ezawateorico2/3}, taking results of \cite{Zheng},  is also qualitatively captured by our analysis made in 
section \ref{sec4}, figure \ref{figdeg}, in the sense that this energy gap goes to zero in the crossover region between spin polarized and unpolarized phases. The increase of level crossings for $\nu=2/3$, as regards $\nu=2$,  
could also have an effect for the appearance of the so called non-QH states, as suggested in \cite{kumada22/3}. 

Anyway, while a more complete microscopic theory might bring new light on the BLQH physics, we think that our present proposal based on composite fermions offers alternative perspectives worth exploring.

\section{Conclusions and outlook}\label{comments}

The physics of multicomponent quantum Hall systems, and particularly the bilayer case, is very rich. The fractional case incorporates extra ingredients that makes the problem even much more interesting. 
In this article we have analyzed the bilayer (four components) case at fractional values $\nu=2/\lambda$ of the filling factor. We have obtained the phase diagram structure of the balanced case by using 
an overcomplete set of coherent (semiclassical) variational states previously introduced. The Hamiltonian used is an adaptation of the integer case, $\nu=2$, to an arbitrary odd number of magnetic flux quanta $\lambda$ 
per electron, to make it intensive for a formal study of the large $SU(4)$-isospin $\lambda$ limit.  We have also performed a 
numerical diagonalization of the Hamiltonian and compared exact (quantum) with mean-field (semiclassical) results for the ground state. The accordance is quite good in the spin and ppin phases, but not 
in the canted phase, where degeneracies and energy level crossings occur, specially at large $\lambda$. We have also analyzed the full energy spectrum and we have found an energy band arrangement in spin and ppin 
phases. The particular structure of these energy bands has also been analyzed in terms of angular momentum and layer population imbalance quantum numbers. An experimental corroboration of these 
band energy formation would enforce our theoretical work. To finish, just to say that a generalization of the previous 
study to arbitrary $N$ component quantum Hall systems at fractional filling factors $\nu=M/\lambda$ could also be (in principle) carried out with the help of our recent construction of $U(N)/[U(M)\times U(N-M)]$ coherent states 
\cite{APsigma} (this is still work in progress).

\section*{Acknowledgements}

The work was supported by the  Spanish project FIS2014-59386-P (Spanish MINECO and FEDER funds). C. Pe\'on-Nieto acknowledges the research contract with Ref. 4537 
financed by the  project above.

\appendix

\section{Orthonormal basis for arbitrary $\lambda$}\label{basissubsec}

In Refs. \cite{GrassCSBLQH,JPCMenredobicapa} we have provided a Fock space representation of the BLQH basis states \eqref{basisvec} for fractional 
filling factor $\nu=2/\lambda$. The general expression is given by the action of creation operators $\mathbf{a}^\dag$ and $\mathbf{b}^\dag$ in layers 
$a$ and $b$ [see \eqref{calzeta} for the definition of $2\times 2$ matrix annihilation operators  $\mathbf{a}$ and $\mathbf{b}$] acting on the 
Fock vacuum $|0\ra_\mathrm{F}$ as 
\bea
|{}{}_{q_a,q_b}^{j,m}\ra&=&\frac{1}{\sqrt{2j+1}}\sum_{q=-j}^{j}(-1)^{q_a-q}\label{basisinfock2}\\
&\times&\frac{\varphi^{j,m}_{-q,-q_a}(\mathbf{a}^\dag)}{\sqrt{\frac{\lambda!(\lambda+1)!}{(\lambda-2j-m)!(\lambda+1-m)!}}}
\frac{\varphi^{j,\lambda-2j-m}_{q,q_b}(\mathbf{b}^\dag)}{\sqrt{\frac{\lambda!(\lambda+1)!}{m!(2j+m+1)!}}}
\;|0\ra_\mathrm{F},\nn
\eea
where 
\bea
\varphi_{q_a,q_b}^{j,m}(Z)&=&\sqrt{\frac{2j+1}{\lambda+1}\binom{\lambda+1}{2j+m+1}\binom{\lambda+1}{m}}\label{basisfunc}\\
&\times& \det(Z)^{m}\cD^{j}_{q_a,q_b}(Z),\; \begin{matrix}
2j+m\leq\lambda, \\ q_a,q_b=-j,\dots,j, \end{matrix}\nn\eea
are homogeneous polynomials of degree $2j+2m$ in four complex variables arranged in a $2\times 2$ complex matrix $Z=(z_{kl})$. Here  
\bea
&& \cD^{j}_{q_a,q_b}(Z)=\sqrt{\frac{(j+q_a)!(j-q_a)!}{(j+q_b)!(j-q_b)!}}
 \sum_{k=\max(0,q_a+q_b)}^{\min(j+q_a,j+q_b)}\label{Wignerf}\\ 
&& \binom{j+q_b}{k}\binom{j-q_b}{k-q_a-q_b}   z_{11}^k
z_{12}^{j+q_a-k}z_{21}^{j+q_b-k}z_{22}^{k-q_a-q_b},\nn\eea
denotes the usual Wigner $\cD$-matrix \cite{Louck3} with angular momentum $j$. The set  of polynomials \eqref{basisfunc} verifies the
closure relation
\begin{equation}\sum^{\lambda}_{m=0}\!\!\sum_{j=0;\um}^{(\lambda-m)/2}\!\!\sum^{j}_{q_a,q_b=-j}\!\!
\overline{\varphi_{q_a,q_b}^{j,m}({Z'})}\varphi_{q_a,q_b}^{j,m}(Z)=K_\lambda(Z'^\dag, 
Z),\nn\end{equation}
with $K_\lambda(Z'^\dag,Z)=\det(\sigma_0+Z'^\dag Z)^\lambda$ the so called Bergmann kernel.

\section{Coherent states on $\mathbb{G}^4_{2}$}\label{subsecCS}

An overcomplete set of coherent states for $\nu=2/\lambda$ has been worked out in Ref. \cite{GrassCSBLQH}. 
Coherent states $|Z\rangle$ are labeled  by a $2\times 2$ complex matrix $Z$ (a point on the Grassmannian $\mathbb{G}^4_{2}$)  
and  can be expanded in terms of the orthonormal basis vectors \eqref{basisinfock2} as
\begin{equation}
|Z\ra=\frac{\sum^{\lambda}_{m=0}\sum_{j=0;\um}^{(\lambda-m)/2}\sum^{j}_{q_a,q_b=-j}\varphi_{q_a,q_b}^{j,m}(Z)
|{}{}_{q_a,q_b}^{j,m}\ra}{\det(\sigma_0+Z^\dag Z)^{\lambda/2}},\label{u4cs}
\end{equation}
with coefficients $\varphi_{q_a,q_b}^{j,m}(Z)$ in \eqref{basisfunc}. They can also be written in the form of a boson condensate as (see \cite{GrassCSBLQH}) 
\begin{equation}
|Z\ra=\frac{1}{\lambda!\sqrt{\lambda+1}}\left(\frac{\det(\check{\mathbf{b}}^\dag+
Z^t\check{\mathbf{a}}^\dag)}{\sqrt{\det(\sigma_0+Z^\dag Z)}}\right)^\lambda|0\ra_\mathrm{F},
\label{u4csfock}
\end{equation}
where $\check{\mathbf{a}}^\dag=\um\eta^{\mu\nu}\tr(\sigma_\mu\mathbf{a}^\dag)\sigma_\nu$ denotes the ``parity reversed'' 
$2\times 2$-matrix creation operator 
of $\mathbf{a}^\dag$ in layer $a$  (similar for layer $b$) [we are using Einstein summation convention with Minkowskian
metric $\eta_{\mu\nu}=\mathrm{diag}(1,-1,-1,-1)$]. Coherent states are normalized, $\la Z|Z\ra=1$, 
but they do not constitute an orthogonal set since they have a non-zero (in general) overlap
given by
\begin{equation}
\la Z'|Z\ra=\frac{K_\lambda(Z'^\dag, Z)}{K_{\lambda/2}(Z'^\dag, Z')K_{\lambda/2}(Z^\dag, Z)}\label{u4csov}
\end{equation}
Sometimes it is useful to use a coherent state picture (Bargmann-Fock representation) of a general state $|\psi\ra$ given by 
$\Psi(Z)\equiv K_{\lambda/2}(Z,Z^\dag)\langle Z|\psi\rangle$. For example, the Bargmann-Fock representation of the basis states 
$|{}{}_{q_a,q_b}^{j,m}\ra$ is given by the homogeneous polynomials $\varphi_{q_a,q_b}^{j,m}(Z)$ in in \eqref{basisfunc}. Given a $U(4)$ 
group element (written in block matrix form) \[U=\left(\ba{cc} A& B\\ C &D\ea\right), \quad A,B,C,B\in\mathrm{Mat}(2,\mathbb{C}),\]  a point $Z$ in the 
Grassmannian $\mathbb{G}^4_{2}=U(4)/U(2)^2$ can be identified with $Z=BD^{-1}$ in the chart where $D$ is invertible. From the composition law of two group 
elements $U''=U'U$ we get the (M\"obius-like) transformation $Z'=B''D''^{-1}=(A' Z+B')(C'Z+D')^{-1}$ of $Z$ under a group translation $U'$. 
This transformation also defines a representation of the $U(4)$ infinitesimal generators $\tau_{\mu\nu}$ on the space of holomorphic functions $\Psi(Z)$, 
given in terms of differential operators $\mathcal{T}_{\mu\nu}$ in four complex coordinates $z^\mu=\tr(Z\sigma_\mu)/2, \mu=0,1,2,3$. For example, 
it is easy to see that the differential realization of the imbalance ppin generator $\tau_{k0}/2$ is given by 
$\mathcal{P}_3=z^\mu\partial_\mu-\lambda$, where we use the Einstein summation convention and denote $\partial_\mu={\partial}/{\partial z^\mu}$ and 
$z_\nu=\eta_{\nu\mu}z^\mu$, with $\eta_{\nu\mu}=\mathrm{diag}(1,-1,-1,-1)$ the Minkowskian metric. In addition, 
spin $\mathcal{S}_k$ and $\mathcal{R}_{k3}$ are written in terms of 
$\mathcal{M}_{\mu\nu}=z_\mu \partial_\nu-z_\nu \partial_\mu$ as $\mathcal{S}_i=\frac{\ic}{2}\epsilon^{ikl}\mathcal{M}_{kl}$ and  
$\mathcal{R}_{k3}=M_{k0}$, respectively, where $\epsilon^{ikl}$ is the totally antisymmetric tensor 
(see \cite{GrassCSBLQH,JPA48} for the remainder $\mathcal{T}_{\mu\nu}$ operators). 
With this differential realization, the (cumbersome) computation of expectation values of operators in a coherent state (usually related to 
order parameters) is reduced to the (easy) calculation of derivatives of the Bergmann kernel as: 
\begin{equation}\la Z|T_{\mu\nu}|Z\ra=K_{\lambda}^{-1}(Z,Z^\dag)\mathcal{T}_{\mu\nu}K_{\lambda}(Z,Z^\dag).\end{equation}
We have used this simple formula to compute the expectation values \eqref{expectv}.

To finish this appendix, let us introduce a parametrization of $Z$ in terms of eight angles 
$\theta_{a,b},\vartheta_\pm\in[0,\pi)$ and $\phi_{a,b},\beta_{\pm}\in[0,2\pi)$, given by the following decomposition
\bea Z=V_a \begin{pmatrix}\xi_+  &0\\ 0& \xi_- \end{pmatrix} V_b^\dag,\; \xi_\pm=\tan\frac{\vartheta_\pm}{2}e^{\ic\beta_\pm},\nn\\ 
V_\ell=\begin{pmatrix}\cos\frac{\theta_\ell}{2}&-\sin\frac{\theta_\ell}{2}e^{\ic\phi_\ell}\\ \sin\frac{\theta_\ell}{2}e^{-\ic\phi_\ell} & \cos\frac{\theta_\ell}{2} 
\end{pmatrix},\, \ell=a,b, \label{paramang}
\eea
where $V_{a,b}$ represent rotations in layers $\ell=a,b$ (note their ``conjugated'' character). This parametrization of $Z$ has been useful 
to minimize the energy surface $\la Z|H_\lambda|Z\ra$ for the Hamiltonian \eqref{Hamlambda}.


\end{document}